\documentstyle[12pt, graphics]{article}

\title{Radion Potential and Brane Dynamics}
\author{L. Mersini\\Department of Physics\\University of Wisconsin-Milwaukee\\Milwaukee,
WI 53201\\ Wisc-Milw-99Th-15}
\date{}

\begin{document}
\maketitle
\begin{abstract}
\footnotesize
We examine the cosmology of Randall-Sundrum model in a dynamic setting where scalar fields are present in the bulk as well as the branes. This generates a mechanism similar to that of Goldberger-Wise for radion stabilization and the recovery of late-time cosmology features on the branes. Due to the induced radion dynamics, the inflating branes roll towards the minimum of the radion potential, thereby exiting inflation and reheating the Universe. In the slow roll part of the potential, the TeV branes have maximum inflation rate and energy as their coupling to the radion and bulk modes have minimum suppresion. Hence, when rolling down the steep end of the potential towards the stable point, the radion field (which appears as the inflaton of the effective 4D theory in the branes) decays very fast and reheats the Universe. This process results in a decrease of the brane`s canonical vacuum energy, $\Lambda_4$. However, at the minimum of the potential $\Lambda_4$ is small but not neccessarily zero and the fine-tuning issue remains. Density perturbation constraints introduce an upper bound on $\Lambda_4$. Due to the large radion mass and strong suppression to the bulk modes, moduli problems and bulk reheating do not occur. The reheat temperature and a sufficient number of e-folding constraints for the brane-universe are also satisfied. The model therefore recovers the radiation dominated FRW universe. 
\end{abstract}
\pagebreak

\section{Introduction.}
\setcounter{equation}{0} The suggestion of lowering the
fundamental scale of quantum gravity to as low as TeV offers a promising scenario in addressing important cosmological
issues. Recently Randall and Sundrum (herein refered to as the RS model) [1] demonstrated that can be achieved without the need for large extra dimensions but instead
through curved 5-dimensional space-time $(AdS_5)$ that generates
an exponential suppression of scales at the boundary where the TeV
brane is located.  

Brane cosmology of the RS model deviates
from the usual FRW Universe. Friedman equation
for brane expansion produces an unconventional relation for
the Hubble constant: $H \sim \rho$ instead of $H \sim
\sqrt{\rho}\;$ [10].  Various authors suggested ways of
obtaining the usual relation $H \sim \sqrt{\rho}$ for
cosmology. However many phenomenological and
astrophysical constraints like inflation and the moduli problem,
the cosmological constant, BBN, reheat temperature $T_{RH}$, sufficient
e-folding, density perturbation and large scale structure, etc. make
this a formidable task. The original RS model was extended by I.
Oda [2] to include many static branes, while T. Nikei [3]
found solutions in the case of two branes inflating in their transverse space.

Based on the RS model, the framework of this paper is a generalized scenario of {\em many branes that are inflating in their transverse space}. TeV branes can have positive as well as negative tensions.  In Sec.2 we review the exact solution in closed form
of [6]. A useful result of [6] is that each brane expands according to its canonical,
effective expansion rate $H_{\mbox{ieff}}$, that depends on the
position of the brane in the fifth dimension. The 5th dimension is required to remain static during and after inflation.

In Sec.3 we allow for matter in bulk and consider rolling branes. In the effective $4D$ picture, the energy of the brane is replaced with a slow-roll potential
(inflation) which comes to a steep end near the minimum,
resulting in reheating. As expected, this requires a potential for stabilizing the radion
during and after inflation. Radion potential induces
inflation in the branes and decays when branes
roll to the stable minimum of the potential.  This confirms Dvali and Tye's
view [4] that although inflaton is a brane mode in the ground
state it behaves as an ``inter-brane'' mode that describes the
relative separation of branes in the 5th dimension. It is this
relative separation that induces inflation on the branes with the
slow-roll properties. It is assummed that the
inflaton is very weakly coupled to the bulk modes because the radion is heavy. This fact avoids reheating of the extra dimension as well as overclosure of the
universe from the KK excitations. It also preserves the
success of BBN by ensuring that the radion has settled at its
minimum and that the size of the extra dimension has not evolved since.
For this to be the case, two requirements need be satisfied
([5]): the radion stabilized by an effective
potential $V_r$ must have a mass $m_r$ larger than the expansion rate $H$; and the brane must have negligible
thickness otherwise the universe cannot be treated as $4$ dimensional at
distances of $H^{-1}$ i.e. $H^{-1} > L$, where $L$ is the size of the extra dimension.

We consider a potential for the radion that satisfies the above
requirements in Sect.3. and join the slow roll solution of [6] reviewed in Sect.2  to the solution found in [7] that is valid during reheating, when branes reach the steep end in the potential
.The number of e-foldings and the
reheating temperature are calculated.

The issue of the effective $4D$ cosmological constant $\Lambda_4$ and
a summary of results are presented in Sec.4. The radion potential induces a canonical
cosmological constant on the branes. $\Lambda_4$ is a
dynamic quantity. When branes are displaced away from
the minimum, in the slow-roll region, $\Lambda_4$ results in inflation of the branes.   However when branes roll towards the
 minimum of the bulk potential, the reheating regime, $\Lambda_4$ very fast takes the value $\Lambda_4^{min} = V_{r, min} \simeq 0$ before BBN
time.  That results in a recovery of the usual Friedman equation
for an FRW radiation dominated Universe. Density perturbation constraints put an upper bound on the value of $\Lambda_4^{min} = V_{r min}$. The lower bound $\Lambda_4^{min}=0$ is the familiar fine-tuning problem of the cosmological constant.

\section{Review of the Inflating Branes Model}

Solutions found in [6] for a five dimensional model with many branes along the fifth compact direction $z$, that inflate in their transverse space $x_i$, extend the recent models of RS, I.Oda and T.Nihei [1,2,3]. 

\noindent The 5D metric ansatz is
\begin{equation}
ds^2 = g_{MN} dx^M dx^N = f(z)^2 dt^2 - f(z)^2 v(t)^2 dx^2 - f(z)^2 dz^2
\end{equation}
with $z$ ranging between $0$ and $2 L$ and $M,N$ running over the five dimensions.
 
\noindent The 5D Einstein-Hilbert action is:
\begin{equation}
S = \frac{1}{2\kappa^2_{5}} \int d^4 x \int^{2L}_0 dz \sqrt{-g} (R+2\Lambda_5)
+ \sum^n_{i=1} \int_{z=L_i} d^4 x \sqrt{-g_i}({\cal L}_i + V_i)
\end{equation}
where the 4D metric $g_{i\mu\nu}$ is obtained from the 5D one as follows:
$g_{i\mu\nu}({\bf x},t) = g_{MN}({\bf x},t,z = L_i)$; and ${\cal L}_i ,(V_i)$
is the lagrangian of the matter fields, (the constant vacuum energy),
in the i-th brane. The 5D gravitational coupling constant $\kappa_{5}$ is related to the 4D Newton constant $G_N$ through $\kappa_5 =16 G_{N}\pi \int dz\sqrt{-g_{55}}$ $=16 G_{N}\pi L_{phys} $.

The dynamics of the boundaries with energy density  denoted by ${\cal L}_i$ (where we have separated a constant vacuum energy contribution, $V_{i}$) are based on the slow roll assumption. The following equations result from varying the action in Eqn.(2.2) with
respect to the metric, [1,2,3,6]
\begin{eqnarray}
\frac{1}{f^2}\left[ (\frac{\dot{v}}{v})^2 \right] - \frac{1}{f^2}
\left[\frac{f''}{f} + (\frac{f'}{f})^2 \right] & = &
- \frac{\kappa^2}{3 f}[\sum_i({\cal L}_i + V_i)\delta(z - L_i)] + 
\frac{\Lambda_5}{3}\nonumber \\
& & \nonumber\\
\frac{1}{f^2}\left[ (2\frac{\ddot{v}}{v})+ (\frac{\dot{v}}{v})^2\right] 
- \frac{3}{f^2}\left[ \frac{f''}{f} + (\frac{f'}{f})^2\right] & = &
- \frac{\kappa^2}{f}\left[ \sum_i({\cal L}_i + V_i)\delta(z - L_i)\right]
+{\Lambda_{5}}\nonumber\\
& & \nonumber\\
\frac{1}{f^2}\left[ (\frac{\ddot{v}}{v})+ (\frac{\dot{v}}{v})^2\right]
 - \frac{2}{f^2}(\frac{f'}{f})^2 & = & \frac{\Lambda_5}{3} 
\end{eqnarray}
where prime and dot denote differentiation with respect to z and t
respectively, and $v$ satisfies
\begin{equation}
v(t) = v(0)exp[Ht]
\end{equation} 
Eqns. (2.3) become
\begin{equation}
\begin{array}{lcc}
\frac{H^2}{f^2} - \frac{1}{f^2} \left[ \frac{f''}{f} + 
                  (\frac{f'}{f})^2 \right] & = & 
- \frac{\kappa^2}{3f} [\sum_i({\cal L}_i + V_i)]\delta(z - L_i) + 
\frac{\Lambda_5}{3} \\
& & \\
\frac{H^2}{f^2} - \frac{1}{f^2}(\frac{f'}{f})^2 
 =  \frac{\Lambda_5}{6}
\end{array}
\end{equation}
The effective expansion rate is defined as $H_{eff} = H / f(z)$.   
Each brane,
 depending on its position in the fifth dimension, feels a different 
expansion rate resulting from scaling of the canonical proper time with the position of the brane, $d \tau_{i} = f( L_{i}) dt$. Thus the effective expansion rate for brane-bound observers, of each brane located at $z = L_i$ in the
extra dimension, in their canonical coordinates, is $H_{ieff} = H / f(L_i)$. Denote the absolute value of $\Lambda_5$ by $\Lambda$. Since we consider an  $ADS_5$ 
geometry, $\Lambda$ is positive and satisfies the following relation
\[ \Lambda = - \Lambda_5 .\]
The solutions of Eqn's. (2.5) are, [6]
\begin{eqnarray}
 f(z) &=& \alpha \displaystyle{\sinh[g(z)]^{-1}}                \\
g(z)  &=&\left(\displaystyle{\sum^{n-1}_{i=1}}(-1)^{i+1}|z-L_i|+L\right)(-\beta)\\
g''(z)&=& 2\left(\displaystyle{\sum^{n-1}_{i=1}}(-1)^{i+1} 
            \delta |z - L_i| \right) (-\beta) 
\end{eqnarray}
\vspace*{0.3in}
\begin{center}
{\bf Figure 1.0} 
\end{center}
The even $(n-1)$ branes are located at
$L_i (i = 2, \dots n-1)$ with alternating positive and negative tensions such that the number of positive
energy branes equals the number of negative energy branes (Fig.1.0).\\
The constants in  (2.6-2.8) are given in terms of the parameters of the model by; $\beta = H$ and 
$\alpha^2 = \frac {6 H^2}{\Lambda}$. The value of $ H $ is determined by $\Lambda$ and $ L$.
The junction conditions, with a static fifth dimension, on the energy densities ${\cal L}_i$ of each brane, and their canonical expansion rates $H_{ieff}$ ,(Fig.2.0), become
\begin{eqnarray}
{\cal L}_i +V_i   &=& \frac {2}{\kappa_5} (-1)^{(i+1)} \sqrt{ \frac {\Lambda}{6} }(\displaystyle \cosh[g(L_i)]) \\
H_{ieff} &=& \sqrt{ \frac {\Lambda}{6} } (\displaystyle  \sinh[g(L_i)])
\end{eqnarray}   
\vspace*{0.3in}
\begin{center}
{\bf Fig. 2.0 } 
\end{center}
The solution for the warp factor $f(z)$, Eqns.(2.6-2.7), Fig.1.0, has a minimum at $z=0, 2L$ and a maximum at $z= L$ \footnote{For subtleties related to ensuring the periodicity of $f(z)$ and the overlap of branes located at $z=0, 2L$ see [2].}. The physical length, $L_{phys}$, depends on the energies of the branes because they are gravitational sources for the curvature of the fifth dimension. 
It is clear from Eqns.(2.9, 2.10) that branes positioned near the minimum of the warp factor, $z=0, 2L$ (the 'TeV branes') in the fifth dimension, have the maximum energy and effective expansion rate allocated to them. Branes near the position of the  maximum warp factor $z=L$ ('Planck branes'), are almost devoid of energy and have the lowest canonical expansion rate \footnote{$H_{ieff}$  at $z=L$ is almost but not identically zero ( Fig.2.0).Similar result have also been obtained by [9,11]}.
Junction conditions of Eqn.(2.9,2.10) give the following Friedmann equation for the expansion rate and the energy density for the branes
\begin{equation}
|\frac {\kappa_5 ( {\cal L}_i + V_i)}{2}|^2 - H_{ieff}^2 = \frac {\Lambda}{6}
\end{equation}
Notice the unusual 'Friedman relation' between the energy density and the expansion rate for each individual  brane [8,9,10]. An important result of the Eqn. (2.11) is that {\rm each brane has a different expansion rate depending on its position in $S^1$} and that the TeV branes do not have to be negative energy branes in this model
\footnote{This issue was one of the main drawbacks in the original RS model with two branes.}.

\section{Radion Potential and Brane Dynamics}
\setcounter{equation}{0}
{\bf 3.1} {\it Radion potential}

It was noted in [7-12] that the $5D$ metric of Eqn.(2.1)  has a gauge freedom in the ($t, z$) plane. We use this property in what follows.
Consider having a bulk field $\Psi(z,t,x)$ with mass $M$, which is pinned down to the branes through an interaction term   
  ${\cal L}_{int} = (\Psi ^2 - E_i^{2})^2 \delta (z - L_i)$ [11]. The wavefunction is taken to be $\Psi(z,t,x) = \psi(z) \cdot \varphi(t,x)$
and $E_i$ are the vev eigenvalues of the bulk field at the brane location $z=L_i$. By varying the
  5-dimensional action $S_5$ with $z$, and assuming that the values of the field $vev`s$ on the branes, $E_i$, are large, one gets the following equation of motion for the scalar field $\Psi(z,t,x)$ \footnote {$M$ can be as high as $M_{pl}$, hence there is a lower bound of  $10^2 - 10^3$  GeV to it from inflation, [5,17]}
 
\begin{equation}
   \left\{ \partial^2_z + [M^2f^2(z) - E_i \delta (z-L_i)] \right\} \psi(z) = 0\,.
\end{equation}
     
The conformal factor multiplies the $M^2$-term due to $z$ translation
invariance of the metric. The induced effective energy density  for the branes [4,5] is
\begin{equation} 
V^{brane}_{i,eff} = \frac{\sigma_i \dot{L}^2_i}{2} + U(L_i) + \sigma_i 
\end{equation}
where $U(L_i)$, obtained from the interaction term in the lagrangian, ${\cal L}_{int}$, is the induced potential to the $z = L_i$ brane from radion coupling, and $\sigma_i$ is the trace of the stress-energy tensor of the brane \footnote {Assuming the brane tension is large, we will approximate the trace of the stress energy tensor $\sigma_i$ by brane`s tension}.

The kinetic term in the effective lagrangian ( dilaton-type
coupling, first term in Eqn.(3.2)) comes from varying the brane action with respect to its
position $L_i$ relative to other branes and describes the relative motion of the i$^{th}$-brane
with respect to the other branes ([4,5]).

The effective radion potential $V_{r,eff}$ is the sum of two contributions, the energy of the bulk field ($U_{bulk} = \frac{1}{2}(\dot\Psi ^2 + M^2 \Psi^2)$) and the energy of pointlike brane sources, $E_i$ and $\sigma_i$, which are the branes energy densities weighed by the conformal factor.  
In a very interesting paper, authors of the first reference in [5] explicitly show the 'analogue' of a Gauss law for stabilizing the radion, namely;  in order to have a
static fifth dimension, consistent with Einstein
equations and junction conditions near the branes\footnote{ Branes break the $\tau$ translation invariance. Matter and radiation energy densities of the branes are related to first-order metric fluctuations near the branes through junction conditions for the energy density $\rho$ and pressure $p$, $\delta a \simeq \rho$, $\delta f \simeq (2\rho +3 p)$, treated in detail in [5,7,9,10,11].}, the effective potential of the radion should be zero ,$V_{r,eff} = 0$. After putting the bulk field solution back into the action, [11], and integrating out the fifth dimension, the effective radion potential becomes $U(z) =\sum_i U(L_i)$, for the $AdS_5$ patch with the slow-roll regime, where the conditions $\sigma {\dot L}^{2}_{i} , \frac{M_{pl} {U'} }{U} << 1,   \frac{M_{pl}^{2} {U''} }{U} << 1 $ are satisfied.
The 4-dimensional potential induced to the $L_i$-brane, $U(L_i)$ is
\begin{equation}
 U( L_i) = \frac{1}{4}  \sum_i \left[ E_i  - M^{2}
f(L_i)\right] 
\end{equation}


The energy density on the
 $i^{th}$-brane is given by Eqn.(3.2).

\vspace*{0.3in}
\centerline{\bf Figure 3.0}
 The $U(z)$ potential has the slow-roll regime for the most part
  except around $z \approx L$ (Fig.3.0) where it becomes very steep. We can therefore use the
previously found solution of Ref.[6], Eqn.(2.11) for the
  slow-roll part in which $U(L_i)$ is nearly a constant.
\begin{equation}
\frac{\kappa_{5}^2}{2}(U(L_i) + \sigma_i)^2 - H^2_{i,eff} = \frac{\Lambda}{6}\,.
\end{equation}
 The region $ L_n \leq z \leq L$ where the slow roll conditions,
 $ \dot{L}^2_i \sigma < U_i + \sigma_i$ and $ M_{pl}
\frac{U'_i}{U_i}, \frac{M_{pl}^{2} U''_i}{U_i} << 1$, break down is studied under the sudden approximation.
The point $L_n$, where the slow roll regime ends, is taken to correspond to the last 60
e-foldings of the slow-rolling brane.  Let us promote the location of the brane $L_i$ to a scalar field, $\phi_i = M^2_{pl} \cdot L_i$. With this notation, we can calculate the e-foldings and
find the location $L_n$ of the last one, $n = 60$, as follows
\begin{equation}
n \geq 60 = \frac{1}{M^2_{pl}}
\int^{ \phi_n(z=L_n)}_{\phi^{(z=L)}_{min}}
 d \phi \frac{U}{U'} =
\frac{\beta \ell n}{M^2_{pl} }
 \left[ \frac{ch[g(\phi_n)]}{ch[g(\phi_L)]}\right]  
\end{equation}
\footnote{Roughly, the point $L_n$ where the slow roll ends and the brane rolls down the steep part of the potential is given by: $60 \beta^{-1} = 60H^{-1} = \frac{\ell n}{M^2_{pl}}
\left[ \frac{ch(g(\phi_n)]}{ch(g \phi_L)}\right] \simeq L_n - L
\left. \right)$}Note that this potential also satisfies the thin brane requirements for the
 effective expansion rate at the end of inflation
 $H^{-1}_{eff}(z=L) >> L_{phys}$ (Eqn. (2.10), Fig. 2.0). We ignore effects of backreaction or quantum corrections on the geometry.
 \\
 {\bf 3.2} {\it Brane Dynamics }

The kinetic term
  $\frac{\sigma_i \dot{L}^{2}_i}{2}$ is ignored  for the
  slow-roll part up to $L_i = L_n$.
The radion dumps most of its energy to the TeV branes which are in
the slow-roll regime, located far from  $z\simeq L_n$ since its coupling has the lowest suppression
to the branes in that regime, Eqn.(2.9,2.10). The highest suppression is in the Planck
brane at $z = L$ because the maximum of $f(z)$ is at $z = L$ and minimum
at the TeV brane $z \simeq 0, 2L$ , Eqn.(2.6), Fig.1.0.

This confirms Dvali and Tye's view in [4] that inflation is an inter-brane
mode, and that the  radion is the inflaton of each brane. The physical reason why TeV brane-bound observers
feel the highest expansion rate as indicated by Eqn. (2.7, 2.10-2.11) is: TeV branes
 get the maximum inflation rate and consequently the highest vacuum energy because most of the radion energy
is dumped in them, while observers bound to the Planck branes have the lowest vacuum energy and
expansion rate $H_{eff}(L)$ as they get very little energy induced by
the radion due to the suppression with $M_{pl}$-scales. Allowing a bulk potential for stabilizing the radion, also induces a potential for the branes that makes them exit inflation and
recover late-time cosmology.  The volcano type potential $U(z)$, which at the brane location appears as $U(L_i) = U(z=L_i)$, Eqn.(3.3), is almost flat for most of the
rolling time up to a critical distance $z = L_n$ where it becomes
very steep and brane inflation ends rapidly (Fig.3.0). In the first approximation [4,5,6,12] the extra radii is frozen
both during and after inflation as the radion is heavy and stabilized by the
potential in Sect. 3.1. The mass of the radion $m_r$ satisfies
\[ m_r \simeq  U''(z=L) > H_{eff}(z=L) = Hsh[g(L)] . \]
At the minimum of radion potential its energy is released into radiation on the branes. The
inflaton remains very weakly coupled to the bulk modes.  This
means only the brane is reheating thus avoiding the problem of
overclosure of the Universe.  The reason is that the mass of
radion $m_r$ is greater than the effective Hubble constant $H_{eff}(z\simeq L)$ during
the last stages of inflation and also the coupling of branes to the bulk modes are at maximum suppression at $z=L$ (the location of the Planck brane).  It was shown in [5] that if
this is the case, the production of KK modes is greatly suppressed.

These requirements are important as they preserve the success of
BBN. The size of extra dimension does not evolve from the time of
nucleosynthesis and any shift in  Newton's constant $G_N$, related to the shift of the radion minima, [5], is negligible.
The slow-roll conditions break at $z \simeq L_n$ and the branes
roll quickly down to the bottom of the potential. For the steep part of the potential, $L_n \leq z \leq L$, 
we use the $sudden$ $approximation$. It is a good approximation as the roll down time $\Delta \tau$ from $L_n$ to $L$ goes as $\Delta \tau \simeq m_r^{-1}$ thus it is very small\footnote{That is why the name volcano type potential}. Authors of [7] found the following solution for the case when the brane near $z=L_n$ is in a mixed state of inflation and radiation 
\begin{equation}
\left( \frac{\dot{a}}{a}\right)^2_{n} = H^2_{eff}(z = L_n) =
\kappa_{5}^4 \sigma^{2}_{n} - \frac{\kappa_{5}^2 \Lambda_5}{6} + 2 \kappa_{5}^4 \sigma_{n} \rho^{n}_{m}
+ \kappa_{5}^4 \rho^{n 2}_{m} + C e^{-4 H_{eff} \tau_n}
\end{equation}
where
\begin{equation}
\left\{ \begin{array}{ll}
 \rho^n_m = \frac{1}{2} \dot{\phi}^2_n + U(\phi_n) + \rho_r \\
\\
 p^n_m = \frac{1}{2} \dot{\phi}^2_n - U(\phi_n)
 + \frac{\rho_{r}}{3}\\ 
 \end{array} \right.
\end{equation}
\begin{equation}
\left\{ \begin{array}{ll}
 \frac{d\rho_r}{d\tau} + 4 H_{eff} \rho_r = \Gamma_\phi \cdot
 \dot{\phi}^2 \\ 
\\
 \ddot{\phi}_n + (3H_{eff} + \Gamma_\phi)\dot{\phi}_n +
 U' (\phi_n) = 0 \\ 
 \end{array}\right.
\end{equation}
\footnote {see [18] and references therein for Eqns.(3.7-3.10)} 
The index $n$ refers to the point $z=L_n$ where slow-roll approximation breaks down and $'min'$ to the point $z=L$ where the brane expansion rate is minimum. $\sigma_n$\footnote{On the assumption that tension dominates the stress energy trace, $\sigma_n$ now denotes brane tension} is the brane tension and $\rho^n _{m}, p^n_{m}$ denote the energy density and pressure for the branes at $z=L_n$. $\rho_r$ is the part of the energy density that decays to radiation, with decay rate $\Gamma_{\phi}$.
It was shown in [5,7,8,9,10,11]) that one recovers the
radiation-dominated Friedman cosmology at $H_{eff}(z=L) \sim
\sqrt{\tau}$ under the condition:
\begin{equation}
\frac{\rho_m^{L}}{\sigma_{L}} << 1 
\end{equation}
The procedure is straightforward; one solves the Eqns. (3.7-3.8) to find  the inflaton $\phi_n(t)$, and radiation energy density $\rho_r$, and
plugs the result in Eqn. (3.6) to get the Hubble expansion rate.

The field $\phi$ during inflation obeys
\begin{equation}
3 H_{eff}\dot{\phi} = -U'(\phi)   
\end{equation}
At $L = L_n$ our solution Eqn. (3.4) for the slow-roll inflation
is joined to the solution of [7], Eqn (3.6) \footnote{see [7] for details of the case of the inflaton oscillating thereby
reheating the universe, i.e, a  radiation-dominated regime near the
minimum of the potential}. Note that the switch from inflation to
radiation happens during a very short time $\Delta \tau \sim m^{-1}_r$ when
the brane rolls down the steep potential. Therefore we use the sudden approximation when joining our solution [6] to that of [7] at the end of the slow roll regime, $z = L_n$. The brane evolves to a radiation dominated Friedmann cosmology when rolling fast towards $z = L$, the stable minimum of the radion. Although the kinetic term becomes important in the intermediate stage during the steep roll from $z = L_n$ to $z = L$ we consider that the inflaton decays very fast, i.e. this regime occurs in a very short time (from the validity of the sudden approximation, the roll time $\Delta \tau \simeq m_r^{-1}$ is very small as $m_r$ is very large) and therefore can be neglected. Instead let`s concentrate only on the two regimes of: slow-roll inflation, just before reaching $L_n$, and the  reheating/radiation dominated regime around and at $z = L$, where the radion stabilizes.

Let us join the slow-roll solution given in Eqn.(3.4), Ref.[6] to the solution found in [7], Eqn.(3.6), near $z = L_n$ with the notation $U(\phi_n) = U_n$ and tension $\sigma_n$
\begin{eqnarray}
H^2_{eff}    & =& \kappa_{5}^2(U_n + \sigma_n)^2 - \frac{\Lambda}{6} =
\kappa_{5}^2[\sigma_n + \rho^n_m]^2 - \frac{\Lambda}{6} +
Ce^{-4H_{eff}\tau_n} \,, \\
 \rho^n_m &=& \frac{1}{2} \dot{\phi_n}^2 + U_n(\phi) + \rho_r\,,
   U_{min}(z=L) = E_{L} - M^2f(z=L)\,,
   \rho^n_m - \rho_{L} = U_n + \rho_r - U_{min} = -C[1 -
  e^{-4H_{eff}\Delta\tau}] \simeq -C \,.
 \end{eqnarray}
 
Joining the two solutions introduces a constraint on the integration constant $C$ (Eqn.3.14)
that depends on the initial conditions [7]. The constraint is $-C = \Delta U = \rho_r + U_n -U_{min}$ and represents the ammount of energy from initial ($L_n$) to final ($L$) state that has been converted to radiation. 
 In order to recover Friedman cosmology around $z = L$ from Eqn.(3.6) with a radiation dominated brane world and the radion stabilized at its minimum
\[H^2_{eff} (z=L) = \kappa_{5}^2 [U_{min} + \sigma]^2 - \frac{\Lambda}{6}\]
we must require [9,10,11,13]:
\begin{equation}
\frac{U_{min}(L)}{\sigma_L} < < 1 
\end{equation}
 i.e
\begin{equation}
 E_n - M^2 f(L) = U_{min} \simeq 0
\end{equation}
with an upper bound such that it satisfies the constraint of the density perturbations (Weinberg's window).
 The induced vacuum energy $\Lambda_4=U_{min}$ on the
brane comes to an almost zero value during radiation - dominated
time and conventional cosmology at late times is recovered.  At
this point $H_{eff}^{Pl}(L)$ reaches also its minimum value (see Fig.2).

Thus when the brane was displaced from this minimum, it had a large vacuum energy $\Lambda_4$ and expansion rate $H_{eff}$ induced from the radion during the slow-roll regime.  However, due to the
dynamics of the brane and radion stability (going to its minimum),
the vacuum energy felt by brane-bound observers $\Lambda_4$ decreased while the brane rolled towards its
minimum. The minimum of the induced potential on the brane $U(z=L)$ is required to be normalized around
$U_{min} \simeq 0$ in order to recover Friedman cosmology, Eqn.(3.15), hence
does not have to be exactly zero.\footnote{Fine-tuning the expansion rate $H_{eff}(L) = 0$, i.e. $U_{min} =0$, through careful cancellation of the brane and bulk vacuum energies, introduces a particle horizon at $z=L$, as discussed in [6,12]. In this case the integration constant $C$ is related to the area of the horizon.}  

The lifetime of radion on the brane is very short and it decays
before the BBN era into radiation, thus cannot serve as a dark-
matter candidate, (details in [15,17]).  However,
the bulk  with mass $M$ is a good candidate,(as first suggested by [5]), because it has a small
annihilation cross-section. This can be shown through the following estimates.  The cross-section $\sigma_M$ goes as  $M^{-2}$ where $M$ is of order $TeV$ or higher and it is stable due to the fact that bulk modes have the highest suppression in coupling to the branes located at $z=L$ where the radion stabilizies, thus the
condition
\begin{equation}
\frac{n_M < \sigma_M |v|>}{H_{eff}} << 1 \;\;\mbox{or}
 \;\;\frac{M}{H_{eff}} \geq 1 
\end{equation}
is satisfied for reheat temperature as low as $T_{RH} = 10^2\;
\mbox{GeV},$ [5,17].

 Note that due to the small size of the
extra dimension and the warp factor, i.e. the relation
\begin{equation}
 M^2_{pl} = M^3_5 \int^{2L}_0 dz f(z) \sqrt{-g_{55}}
\end{equation}
and the lower bound on the radion mass $m_r$,  $m_r \geq 10^2$-GeV ,([5,17]),  the reheat temperature given by  $T_{RH} \sim \sqrt{\Gamma_\phi M_{pl}},
\; \Gamma_\phi \sim \frac{m^3_r}{M^2_5}$ [17] can easily take values $T_{RH}
\geq 10^2$ GeV.
There is also the constraint of density perturbations at the end of inflation.
\begin{eqnarray}
\frac{\delta \rho}{\rho} \sim 10^{-4} \\
\frac{\delta\rho}{\rho} \sim \frac{H}{M_{pl} \in}, \; \in_{60} \sim
M_{pl} \frac{U'}{U}
\end{eqnarray}

In order for this astrophysical constraint to be satisfied it imposes an upper bound on the value of $U_{ min}= U(L)$. This translates into a constraint on the size $L_{phys}$ of the $AdS$ curvature. Through Eqns.(2.10-2.11, 3.19), the constraint Eqn.(3.19) on the energy $U_{min}$, and $H_{eff}$ at the end of inflation are such that
\begin{equation}
H_{eff}(L)^{2}M_{pl}^2 << (10^{-3} eV)^4
\end{equation}

The short lifetime of the radion
 (which is seen by the brane as the effective $4D$ inflaton field) means that it decays very fast and reheats the
 universe very fast.\\
\section{Conclusions}
Phenomenological models must reproduce conventional cosmology features and satisfy astrophysical constraints in order to become sussessful theories.

In this paper we investigate the cosmology of a generalized RS model in a dynamic setting. A bulk field introduces a potential which stabilizies the radion. Branes roll because their relative separation adjustes to the radion settling towards its stable minimum. Meanwhile, the  radion appears as the inflaton in the effective $4D$ description of the brane world. 
It confirms Dvali and Tye's  point of view [4] that although the inflaton is a brane mode in the ground state, it behaves as an 'inter-brane' mode that describes a relative separation of the branes in the extra dimension. This approach clarifies the physics of brane dynamics during inflation. The radion which depends on the inter-brane separation, in the effective 4D description, drives inflation in the branes. The TeV branes have minimum suppression to the radion modes as compared to the 'Planck branes', thus they will have the maximum inflation confirmed by Eqn. (2.6-2.11). The coupling of 'Planck branes' to the radion and bulk modes have maximum suppression, thereby almost zero inflation rate and vacuum energy since the energy dumped into them by the radion is highly suppressed by Planck mass scales $M_{Pl}$. Branes inflate with different canonical expansion rates because their coupling to the radion depends on their position/angle in the $S^1$.

One of the motivations to consider this model with many branes inflating in their transverse space, is that it allows TeV branes to have positive energy and tension, as compared to the case of two branes only where the geometry forces the TeV brane to have negative tension.

The dynamics of the model (rolling of the branes and radion stability) ensure the recovery of coventional cosmology features at late times. The radion decays very fast in the $4D$ brane world, reheats the Universe and drives a decaying cosmological constant while reaching its stable minimum.
Thus, the  cosmological constant issue becomes a  dynamic problem, where the
 branes adjust their positions
 towards a static solution with a stabilized radion. However the final value of the vacuum energy $\Lambda_4$ felt by brane-bound observers, despite  being very small, needs to be tuned to zero. It has an upper bound in order to satisfy the astrophysical constraint of density perturbations, Eqn.(3.19-3.21). This constraint puts a bound on the bulk curvature (Eqn.3.21).
Due to their final  positioning in the warp factor, coupling of the branes to bulk modes are at maximum suppresion after inflation when the radion has stabilized. This avoids the reheating of the bulk. The brane becomes a FRW radiation-dominated universe.
There is no moduli problem because the radion is heavy, $m_r >> H_{eff}(L)$, and decays very fast thus overclosure of the universe and production of dangerous relics do not arise in this model.
Cosmological and astrophysical constraints such as: reheat temperature, sufficient number of e-folding, negligible changes in the constants of nature, are satisfied.
Bulk modes can provide a good candidate for dark matter as they are cold, stable (almost decoupled from the branes) and the conditions of the smallness of their annihilation cross-section and the reheat temperature are satisfied (see Eqns. (3.17-3.21)).

For the most part,the potential is nearly flat at the slow roll regime and ends steeply to a fast decaying,/reheating regime, near the stable minimum, $z=L$ . The sudden approximation employed in solving the equations in the two regimes (i.e. ignoring the intermediate steep part of the fast roll regime, located between the nearly flat part and  the stable minimum, where the kinetic term is important), is thus justified. The time taken to  roll down the steep part of the potential is proportional to the inverse of the radion mass and therefore very small. The backreaction of brane matter on the geometry and their quantum corrections have been ignored.

Acknowledgment:
I would like to thank Prof. L. Parker for our regular discussions.This
work was supported in part by NSF Grant No. Phy-9507740.

{\large{\bf References}}
\begin{itemize}
\item[{1}]  L. Randall and R.Sundrum, 'A large mass hierarchy from a small extra dimension', {\bf Phys.Rev.Lett.83:3370-3373}, $hep-ph/9905221$
\item[{2}]  Ichiro Oda, {\bf Phys.Lett.B480:305-311}, {\em $hep-th/9908104$}; {\bf Phys.Lett.B472:59-66}, {\em $hep-th/9909048$}
\item[{3}]  T.Nihei, {\bf Phys.Lett.B465:81-85}, $hep-ph/9905487$; N.Kaloper, {\bf Phys.Rev.D60:123506}, $hep-th/9905210$
\item[{4}]  G. Dvali and H. Tye, {\bf Phys.Lett.B450:72-82}, $hep-ph/9812483$
\item[{5}]  C. Csaki, M.Graesser, L. Randall and J. Terning, {\bf Phys.Rev.D62:045015}, $hep-ph/9911406$; C. Csaki and Y. Shirman, {\bf Phys.Rev.D61:024008}, $hep-th/9908186$; G. Felder, L. Kofman and A. Linde, {\bf JHEP 0002:027},$hep-ph/9909508$
\item[{6}]  L. Mersini,`Cosmological constant of the inflating branes in the Randall-Sundrum-Oda model`, $hep-ph/9909494$
\item[{7}]  E.E. Flanagan, S-H. Henry Tye and Ira Wasserman, {\bf Phys.Rev.D62:044039}, $hep-ph/9910498$
\item[{8}]    A.Lukas, B.A.Ovrut and D.Waldram, {\bf Phys.Rev.D61:023506}, $hep-th/9902071$
\item[{9}] C.Csaki, M.Graesser,C.Kolda,J.Terning, {\bf Phys.Lett.B462:34-40}, $hep-ph/9906513$; J.Cline, C.Grojean,G.Servant, {\bf Phys.Rev.Lett.83:4245,1999}, $hep-ph/9906523$; J. Cline, {\bf Phys.Rev.D61:023513}, $hep-ph/9904495$
\item[{10}]   P. Binetruy, C.Deffayet, D.Langlois, {\bf Nucl.Phys.B565:269-287}, $hep-th/9905012$,; P.Kanti, I.I Kogan, K.A. Olive and M.Pospelov, {\bf Phys.Lett.B468:31-39}, $hep-ph/9909481$; P. Krauss, {\bf JHEP 9912:011}, $hep-th/9910149$
\item[{11}]  W.D.Goldberger and M.B.Wise, {\bf Phys.Rev.Lett.83:4922-4925}, $hep-ph/9907447$, {\bf Phys.Lett.B475:275-279}, $ hep-ph/9907218$; H.B.Kim and H.D.Kim, {\bf Phys.Rev.D61:0604003}, $hep-th/9909053$; P.Steinhardt, {\bf Phys.Lett.B.462:41-47}, $hep-ph/9907080$; L.Mersini, {\bf Mod.Phys.Lett.A14:2393-2402}, $gr-qc/9906106$  
\item[{12}]  A.Chamblin, S.W.Hawking, H.S.Reall, {\bf Phys.Rev.D61:065007}, $hep-th/9909205$; B.Bajc and G.Gabadadze, {\bf Phys.Lett.B474:282-291}, $hep-th/9912232$; S.M.Carroll, S.Hellerman,M.Trodden, {\bf Phys.Rev.D61:065001}, $hep-th/9905217$
\item[{13}]    N. Arkani-Hamed, S, Dimopoulos, G.Dvali and N.Kaloper, {\bf Phys.Rev.Lett.84:586-589}, $hep-th/9907209$
\item[{14}]   N.Kaloper and A.Linde, {\bf Phys.Rev.D59:043508}, $hep-th/9811141$
\item[{15}]   J. Lykken and L.Randall, {\bf JHEP 0006:014}, $hep-th/9908076$
\item[{16}]   H. Verlinde, {\bf Nucl.Phys.B580:264-274}, $hep-th/9906182$
\item[{17}]  Daniel J.H. Chung, Edward W. Kolb and Antonio Riotto, {\bf Phys.Rev.D59:023501}, $hep-ph/9802238$
\item[{18}]  A.Albrecht, P.J. Steinhardt, M.S. Turner and F.Wilczek, {\bf Phys.Rev.Lett.48} (1982)
\end{itemize}]
\newpage

\enlargethispage*{4000pt}

\resizebox{3.5in}{2.1in}{\includegraphics{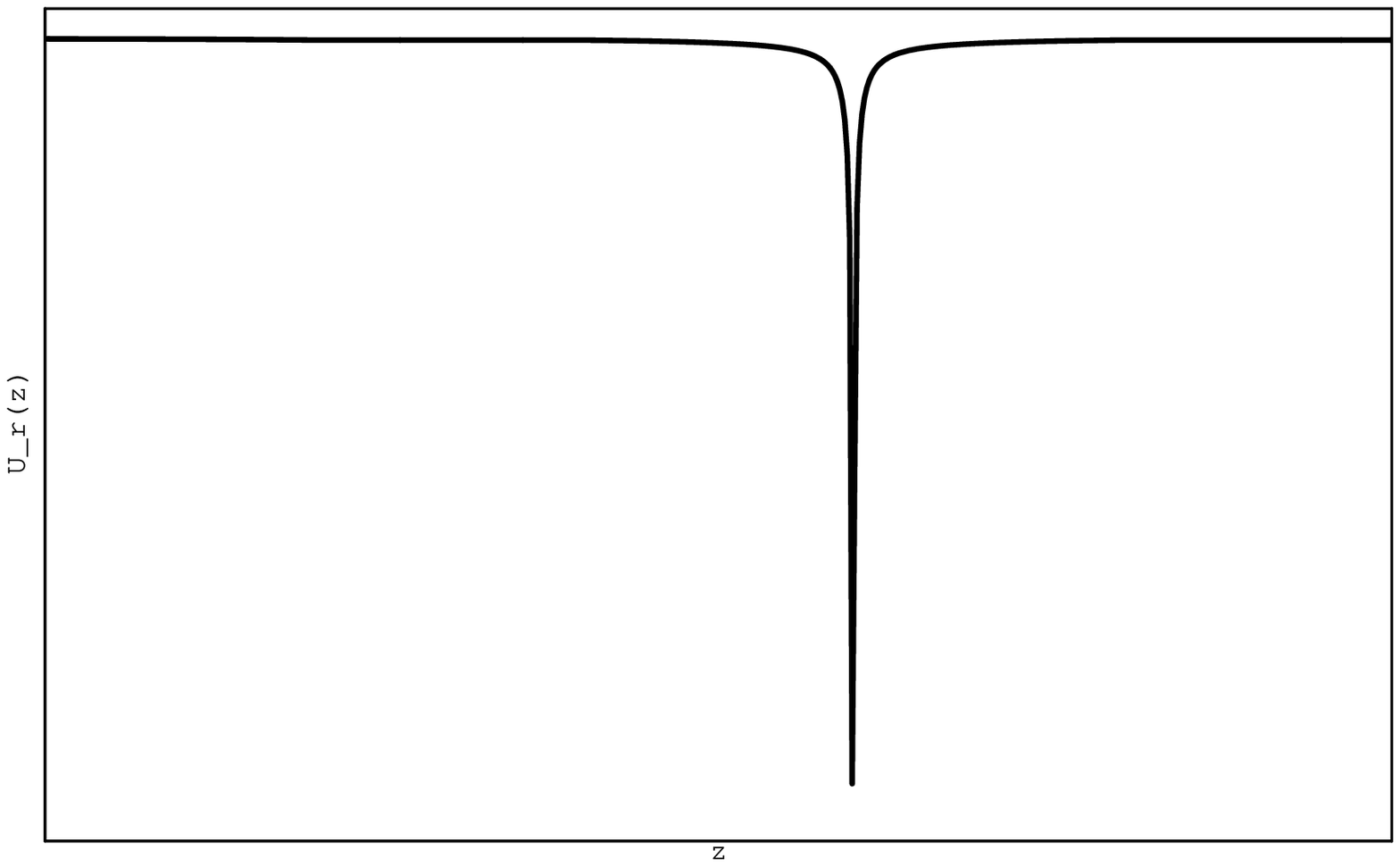}}

\begin{center}
{\bf Fig. 3.0 The radion potential $U(z)$ vs.$z$. TeV branes are located on the flat part (note maximum inflation rate) and roll down the steep end towards the minimum $z=L$. At this point, the radion has stabilizied, branes exit inflation, and reheat the universe} 
\end{center}
\resizebox{3.5in}{2.5in}{\includegraphics{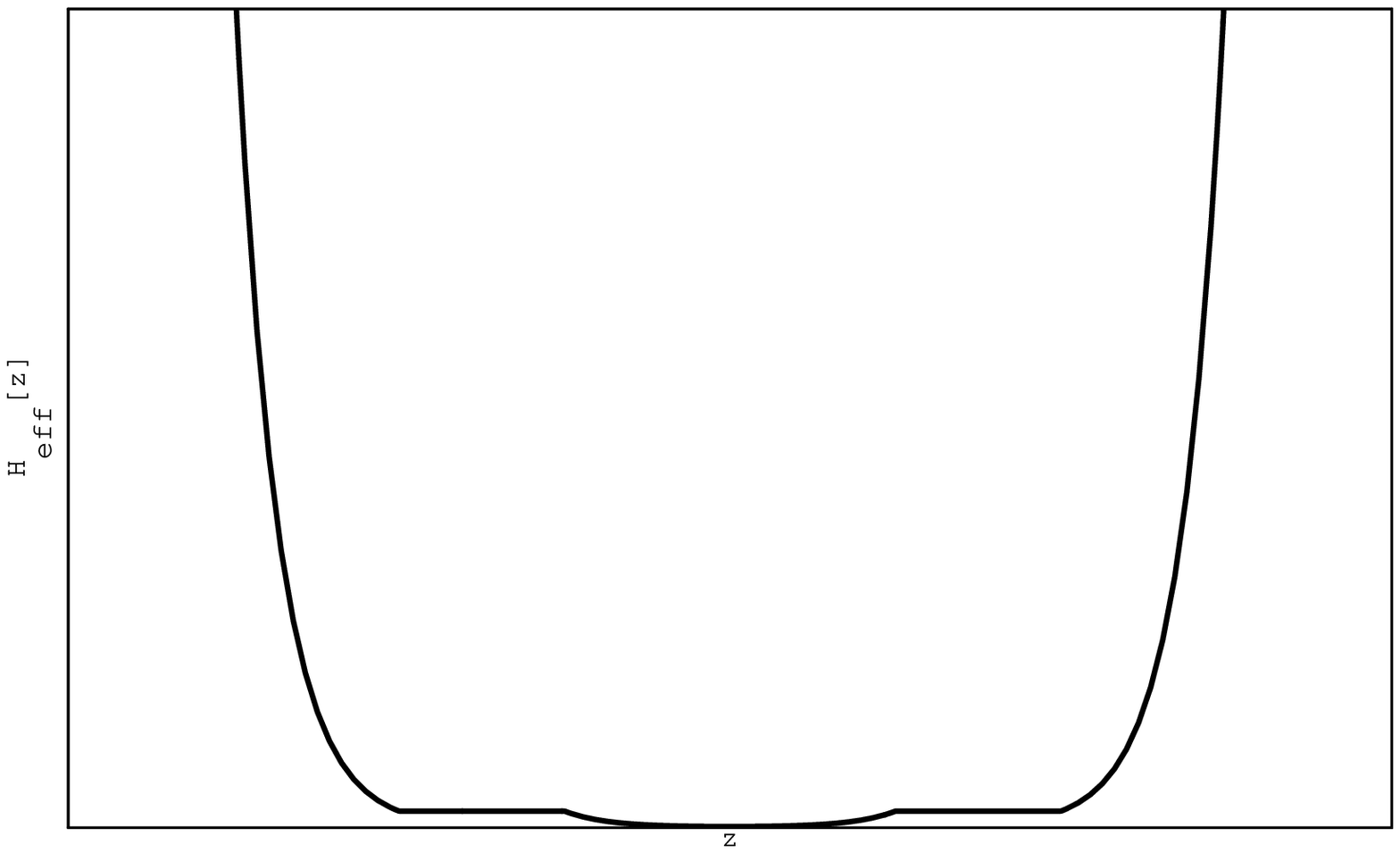}}

\begin{center}
{\bf Fig. 2.0 The effective expansion rate, $H_{eff}(z)$. Each brane located at $z=L_i$ has an expansion rate $H_{eff}(L_i)$} 
\end{center}
\resizebox{3.5in}{2.1in}{\includegraphics{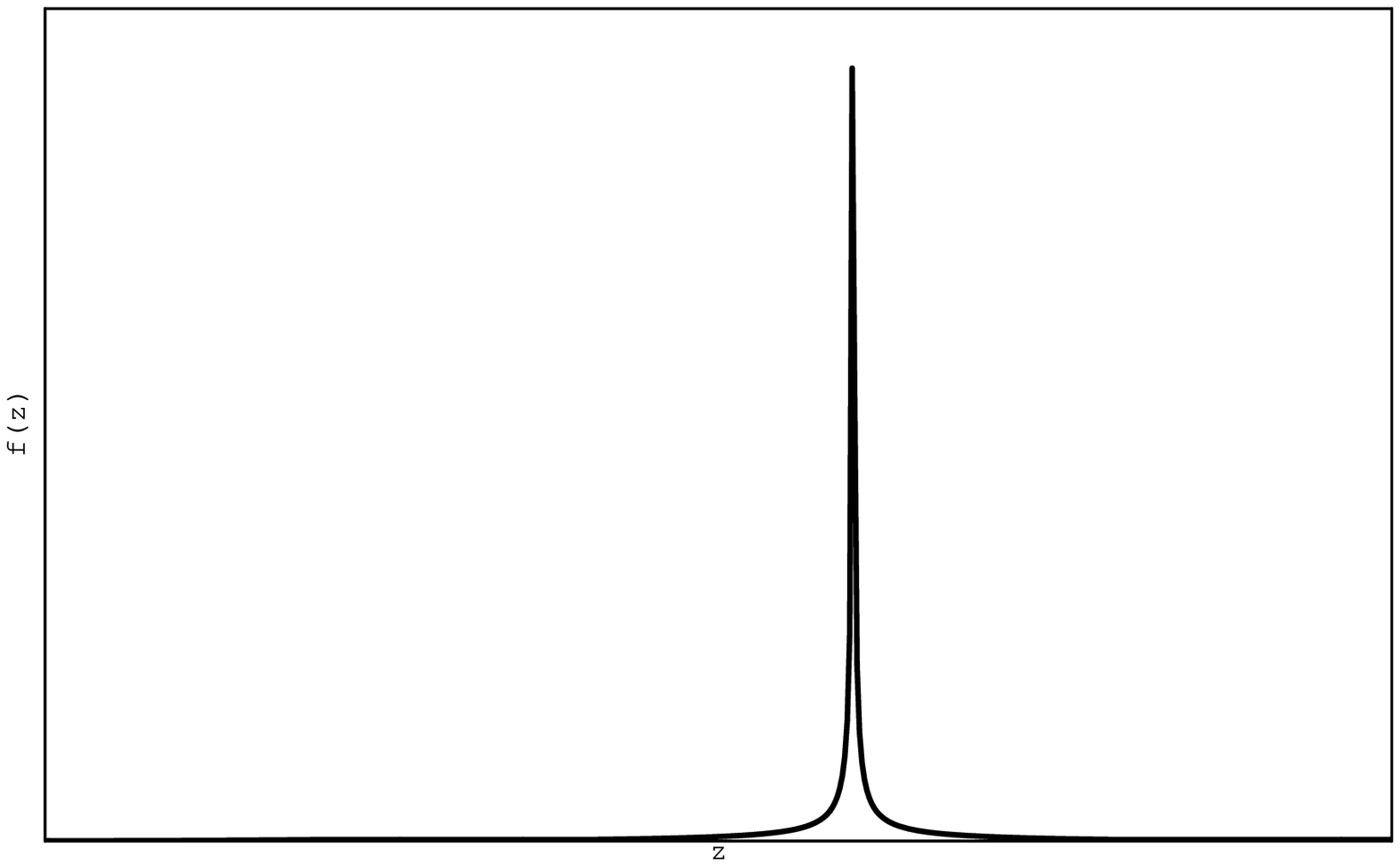}}

\begin{center}
{\bf Fig. 1.0  The scale factor, $f(z)$ vs. $z$ of Sect.2, for the many inflating branes case. Maximum located at $z=L$, minimum at $z=0, 2L$}
\end{center}

\end{document}